## Article information

### Article title
Quasi-DNS with chemical kinetics for a single multiple-injection burner element under stable lean conditions for future gas turbine applications

### Authors
Kazuki Abe*, Youhi Morii, Kaoru Maruta

### Affiliations
Institute of Fluid Science, Tohoku University

### Corresponding author's email address and Twitter handle
kazuki.abe.r2@dc.tohoku.ac.jp

### Related research article
***For a published article:***
K. Abe, Y. Morii, K. Maruta, Proc. 14th Asia-Pacific Conf. Combust., (2023), 224.

### Abstract
We develop a Quasi-DNS with chemical kinetics for a stable lean condition in a single multiple-injection burner element for future gas turbine applications, as a reusable workflow for a coaxial mixing-tube burner. Here, Quasi-DNS denotes a high-resolution, turbulence-model-free calculation; it is not DNS in the strict sense.

The OpenFOAM-based method includes: (i) a simplified 30° sector geometry capturing vortices at the mixing tube outlet, (ii) boundary conditions reproducing coaxial jets and preheated air, (iii) a reduced Correa methane-air mechanism compared with GRI 3.0 using Cantera, (iv) mesh-development and grid-sensitivity assessment based on flame structure and the squared gradient of mixture fraction, and (v) post-processing with methane-based flame index.

Demonstration at an equivalence ratio $\phi = 0.45$ shows limited in-tube mixing and strong outlet vortices, yielding a more extended heat release region than the RANS solution.

- A reusable high-resolution, turbulence-model-free Quasi-DNS workflow with detailed methane-air chemistry is developed for a coaxial mixing-tube burner.
- The workflow combines a reduced methane-air mechanism, refined sector-domain meshes, and scalar-field diagnostics for characterizing mixing and heat release regions.
- The workflow enables analysis of mixing and flame stabilization under stable lean conditions, and provides a high-resolution numerical reference under the stated assumptions for comparison with RANS-based calculations.

## Graphical abstract

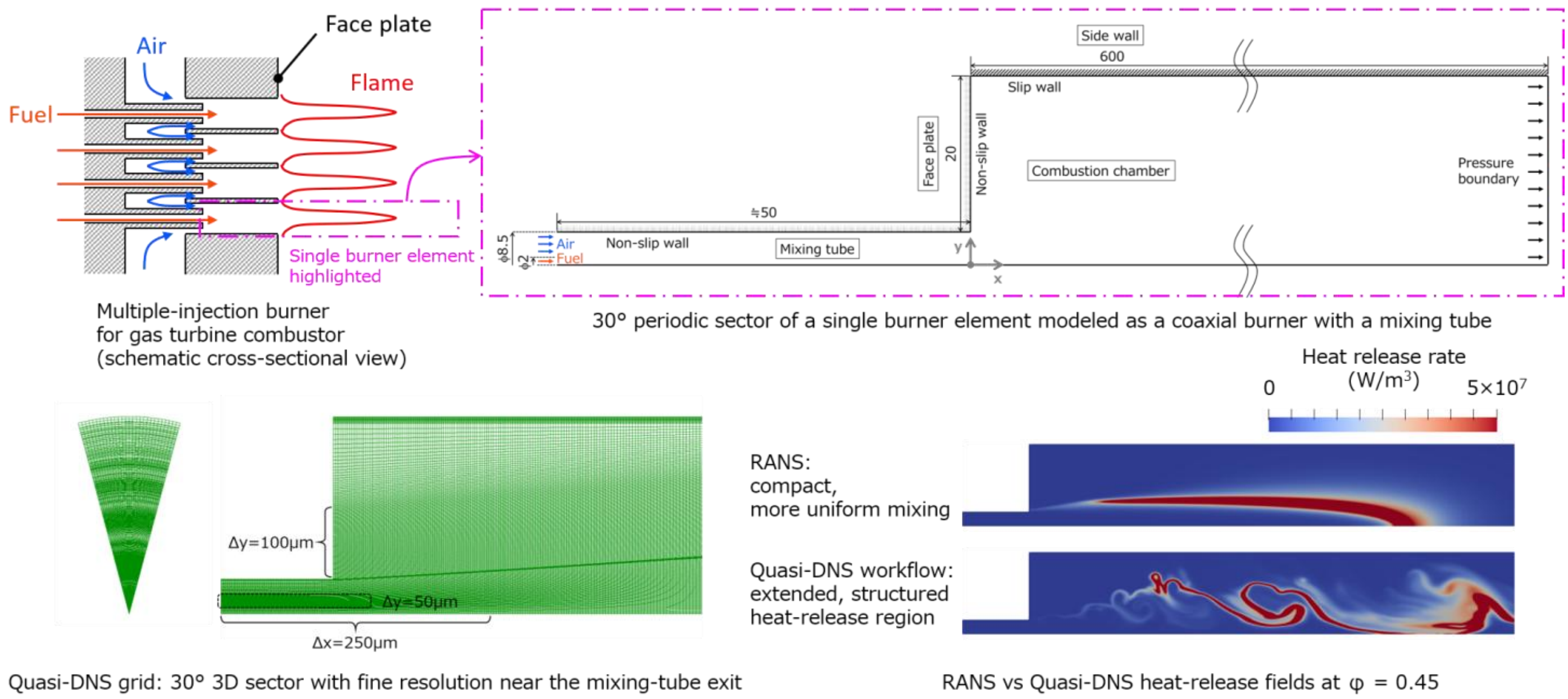

## Specifications table

| | |
|---|---|
| **Subject area** | Engineering |
| **More specific subject area** | Gas turbine combustion and CFD |
| **Name of your method** | Quasi-DNS with chemical kinetics for a single multiple-injection burner element under stable lean conditions for future gas turbine applications |
| **Name and reference of original method** | N/A |
| **Resource availability** | The exact version reviewed in this manuscript is archived on Zenodo (DOI: 10.5281/zenodo.22914633) and linked to the GitHub repository (https://github.com/kazuki-abe-111/quasi-dns-methodsx). The archive includes the OpenFOAM case files, mesh and geometry, initial and boundary conditions, chemical mechanism, solver settings, post-processing scripts, software-version information, and instructions for reproducing the reported diagnostics and figures. |

## Background

Gas turbine combustors for power generation are under strong pressure to achieve higher efficiency and lower emissions while maintaining stable operation over a wide load range [1-2]. In particular, low NOx systems firing natural gas are often operated close to the lean blow-out (LBO) limit, where small changes in mixing or flow conditions can trigger flame extinction [3-4]. The introduction of hydrogen and other low carbon fuels into gas turbine systems promotes lean operation. However, the high laminar burning velocity of hydrogen increases the risk of flashback even under lean conditions [5].

Conventional low NOx combustors for heavy duty gas turbines frequently employ strongly premixed burners with swirled or highly mixed flows [6-7]. These configurations are effective in reducing emissions, but they are susceptible to flashback, especially when hydrogen is blended into the fuel. The multiple-injection burner has been developed as an alternative concept that combines low emissions with high flashback resistance [8]. The multiple-injection burner is already in commercial use for IGCC applications, and further development for natural gas and other fuels is expected [9-10, 41]. In these burners, multiple small coaxial fuel-air jets discharge into mixing tubes where fuel and air are rapidly mixed, and combustion occurs downstream of the mixing region, which helps to achieve both low emissions and resistance to flashback.

The flame behavior in such coaxial burners with mixing tubes is strongly influenced by fuel-air mixing inside the tube and by flow structures at the tube outlet, where the flame is stabilized [11-14]. Previous experimental and numerical studies have examined coaxial jet flames, lifted turbulent $H_2/N_2$ jet flames in hot vitiated coflow, and lifted diffusion flames under laminar or transitional conditions, as well as methane jet flames with hot coflow air [15-18]. However, there are few studies that explicitly consider burners with a mixing tube downstream of the coaxial jet, as adopted in the multiple-injection burner. Since fuel and air are mixed and combusted within a finite residence time in the mixing tube, the mixing process inside the tube can have a strong influence on flame stability.

Prediction of the LBO limit is important for the performance evaluation of gas turbine combustors and element burners [19-20]. In practice, LBO limits are primarily determined by experiments, while numerical simulations are used mainly to support and interpret the measurements. For such purposes, RANS-based approaches have routinely been used, and LES has also been applied in more detailed studies of specific combustor configurations. Although DNS calculations with detailed chemical reactions are computationally expensive, they reproduce experimental results well and are increasingly used to understand turbulent combustion phenomena; they have also been employed in combustion model development [21-26]. In our previous study, we investigated blow-out characteristics of methane flames using transient Quasi-DNS and showed that Quasi-DNS can be useful for understanding blow-out phenomena [27].

The present work focuses on the Quasi-DNS methodology itself and provides a transferable workflow for analyzing flame structure and stabilization in a coaxial burner with a mixing tube under a stable lean operating condition. The method combines a geometry and meshing workflow incorporating sector-domain modeling and mesh resolution assessment, the Correa methane-air reaction mechanism evaluated through comparisons with GRI-Mech 3.0 and relevant published experiments, and scalar-field diagnostics—including the squared mixture fraction gradient and flame index—to characterize mixing and combustion modes near the mixing tube outlet. The present workflow is demonstrated at an equivalence ratio of $\phi = 0.45$. It is intended to provide a high-resolution framework for analyzing mixing and flame stabilization in single multiple-injection burner elements; extinction limits, transient lean blow-out events, and the predictive capability of the method for LBO are outside the scope of this study.

## Method details

### Overall approach

The workflow consists of the following five steps:

Step 1: Define the single element geometry and operating conditions for the coaxial burner with mixing tube.

Step 2: Prescribe inlet and outlet boundary conditions and inlet velocity profiles.

Step 3: Select and validate an efficient methane-air reaction mechanism using Cantera laminar flame speed, ignition delay and counterflow diffusion flame calculations.

Step 4: Construct RANS and Quasi-DNS grids, perform mesh resolution and grid-sensitivity assessments based on flame structure and the squared mixture fraction gradient, and choose numerical settings.

Step 5: Compute diagnostics, including the squared mixture fraction gradient and the flame index, and compare the Quasi-DNS results with the corresponding RANS predictions near the mixing tube outlet.

This work employs a transient Quasi-DNS approach with detailed chemistry to analyze methane-air combustion in a coaxial burner with a mixing tube and downstream combustion chamber. The simulations are performed using the reactingFoam solver in OpenFOAM v1706. The physical and chemical modeling framework follows that used in the authors' previous Quasi-DNS study of the same burner configuration, with the modifications and additional diagnostics introduced in the present work described below. The workflow consists of defining a representative single burner geometry and operating conditions, prescribing inlet and outlet boundary conditions, selecting and assessing a reduced methane-air reaction mechanism, constructing RANS and Quasi-DNS grids with mesh resolution and grid-sensitivity assessments, and applying scalar diagnostics—including the squared mixture fraction gradient and the flame index—to characterize mixing and combustion near the mixing tube outlet.

### Geometry and target conditions

The target configuration represents a single burner element of a multiple-injection burner used in gas turbine combustors, modeled here as a coaxial burner with a mixing tube and a downstream combustion chamber. A schematic of the computational domain is shown in Fig. 1 and the main dimensions and operating conditions are summarized in Table 1. The origin of the coordinate system is located at the center of the coaxial inlet plane, with x along the burner axis and θ in the azimuthal direction of the 30° sector. At the inlet, a coaxial nozzle supplies methane at the center and air in an annulus. The inner fuel port is a circular tube with an inner diameter of ϕ2 mm, and the air is supplied through a concentric annular passage with an outer diameter of ϕ8.5 mm at the inlet plane.

In the practical multiple-injection burner, the fuel and air passages are separated by a tube wall with a finite thickness of about 1 mm. Resolving the recirculation that forms behind this edge would require a much finer grid and significantly increase the computational cost. In the present model, the fuel and air boundaries are therefore represented as baffle boundaries with zero thickness. This simplification removes the small scale recirculation behind a thick lip and allows the analysis to focus on the shear mixing layer between the fuel and air streams, which is expected to play a dominant role in the flame stabilization mechanism near the mixing tube outlet. The present single element model is intended to resolve the local mixing and flame-stabilization processes near an individual mixing tube outlet, rather than to reproduce interactions among multiple burner elements in a full combustor.

Downstream of the inlet, a straight circular mixing tube with an inner diameter of ϕ8.5 mm is attached. Its axial length is approximately 50 mm, providing a finite residence time for fuel-air mixing before the mixture enters the combustion chamber. The mixing tube discharges into a cylindrical combustion chamber with an inner diameter of ϕ48.5 mm and an axial length of 600 mm. The chamber length is chosen so that the outlet boundary does not influence the flame stabilized near the mixing tube exit while keeping the computational cost acceptable.

Methane, representing the main component of natural gas, is used as the fuel. The overall equivalence ratio $\phi$ at the central nozzle is varied based on previous element burner tests and preliminary simulations. In this method article, the Quasi-DNS workflow is demonstrated primarily for a partially premixed flame at $\phi = 0.45$, which corresponds to a stable lean condition relevant to multiple-injection burner operation.

## Boundary conditions and inlet velocity profiles

The inlet boundary conditions are defined to represent fully developed internal flows at the nozzle exits. Methane is supplied through the inner circular port and preheated air through the annular port. For the fuel inlet, a Poiseuille velocity profile in a circular tube is prescribed, assuming a fully developed laminar flow. The fuel temperature is 25 °C and the composition is methane, with the equivalence ratio determined by the target operating condition. For the air inlet, a Poiseuille-type velocity profile corresponding to a double circular tube is used to represent the annular passage. The air temperature is 400 °C, and the composition is dry air at atmospheric pressure, typical of gas turbine combustor inlet conditions.

The Reynolds numbers based on the bulk velocities and hydraulic diameters are approximately $2.2\times10^3$ on the fuel side and $5.5\times10^3$ on the air side. These values are of the same order as the critical Reynolds number for transition in a circular tube ($\simeq 2.3\times10^3$), and it is not clear whether the inlet flows are laminar or turbulent in practice, as this depends on the level of upstream disturbances [28]. In addition, strong shear at the fuel nozzle outlet, which is located just downstream of the inlet boundary in this model, rapidly modifies the velocity profiles. For these reasons, prescribing laminar Poiseuille profiles for both fuel and air is considered a reasonable approximation; differences between laminar and turbulent inlet profiles are not expected to have a significant effect on the flame stabilization region in this configuration.

At the downstream end of the combustion chamber, a pressure outlet boundary condition is applied with fixed static pressure and no reverse flow. The walls of the mixing tube and the face plate are treated as no-slip boundaries, the side wall of the combustion chamber is treated as a slip boundary, and all walls are adiabatic. The two azimuthal faces of the 30° sector are treated as periodic boundaries.

## Numerical solver and time integration

All simulations are performed using the reactingFoam solver in OpenFOAM v1706, as in the authors' previous work [27]. The governing equations are the Navier-Stokes equations coupled with conservation equations for energy and species, including detailed chemical reaction source terms. Time integration uses a one-parameter family of schemes that blends Crank-Nicolson and implicit Euler methods via a time integration parameter $\theta$ [29]. In this study, $\theta$ is set to 0.4. Preliminary calculations were conducted for several values of $\theta$, and $\theta = 0.4$ was selected because it provided stable calculations. Spatial discretization employs a second-order central differencing scheme with TVD limitation for divergence terms. Pressure-velocity coupling is handled using the PIMPLE algorithm.

Two types of simulations are carried out on the same geometry:

- Unsteady RANS calculations use the standard k-ε turbulence model. The reaction source terms are solved directly on a relatively fine grid, without additional turbulent combustion models or filtering [31]. The mesh is much finer than in typical engineering RANS applications to allow a consistent comparison with Quasi-DNS.

- Quasi-DNS calculations do not employ any turbulence model. Only the resolved scales are simulated, and no subgrid-scale model is introduced as in LES. The grid is refined to resolve flame structures and key flow features near the mixing tube outlet, and its adequacy is assessed through the mesh development and grid sensitivity procedure described below. The time step is adjusted to keep the Courant number below 0.3, which leads to time step sizes on the order of $10^{-8}$ s for the present Quasi-DNS cases. The Quasi-DNS simulations were initialized from the corresponding RANS solution and were performed on 960 CPU cores, with a computational cost on the order of $10^5$ core-hours.

In the present study, the term “Quasi-DNS” denotes a high-resolution simulation performed without an explicit turbulence model or subgrid scale combustion model, with finite-rate chemistry directly incorporated. The present calculation does not satisfy the strict definition of DNS because all relevant turbulent length and time scales, including the Kolmogorov scale throughout the entire computational domain, are not fully resolved.

The overall numerical framework, including reactingFoam and detailed chemistry, has been validated in previous fundamental work by reproducing the unsteady FREI (flames with repetitive extinction and ignition) phenomenon in a micro flow reactor with a controlled temperature profile [32].

**Reaction mechanism selection and validation**

A large mechanism such as GRI 3.0 is not suitable for practical Quasi-DNS calculations because of its high computational cost [33]. For the present study, a reduced reaction mechanism with a small number of elementary reactions but with sufficient accuracy under gas turbine conditions is required. The air temperature in the combustor is 400 °C at atmospheric pressure, which is representative of gas turbine combustor inlet conditions. Under such conditions, the Correa mechanism, which was developed for gas turbine combustor applications and has been widely used in combustion studies, is adopted [30, 34-35]. It consists of 16 species and 41 reactions.

The validity of the Correa mechanism is verified using Cantera 2.6.0 [40]. Laminar burning velocities are first calculated at atmospheric pressure for methane-air mixtures with fuel and air both at 25 °C. These preliminary checks show good agreement between the Correa mechanism and GRI 3.0 on the lean side, with larger differences on the rich side. For the inlet conditions used in this study, with air at 400 °C and fuel at 25 °C, detailed comparisons of laminar burning velocity obtained with the two mechanisms and experimental data from Ref. [46] for reference are then performed, as shown in Fig. 2. The agreement at 400 °C is relatively good from lean to rich conditions, indicating that the Correa mechanism is appropriate for the elevated air temperature used here.

Ignition delay times are also compared at atmospheric pressure for several equivalence ratios, including $\phi = 0.4$ and 1.0, over a range of temperatures. Representative results for $\phi = 1.0$, together with experimental data from Ref. [47], are shown in Fig. 3. The differences between GRI 3.0 and the Correa mechanism are small and of the same order of magnitude. Taken together, these comparisons support the use of the Correa mechanism for methane-air combustion under the present conditions.

In the OpenFOAM implementation, the standard reactingFoam solver assumes a Lewis number Le = 1 and does not include species-dependent molecular diffusion coefficients [36-39]. To assess the impact of this approximation, counterflow diffusion flame calculations are carried out using Cantera with both GRI 3.0 and the Correa mechanism, and the results are compared with reactingFoam simulations using the Correa mechanism (Fig. 4). Even in these laminar diffusion flames, where molecular diffusion is expected to have a strong effect, the differences in temperature distribution are small. Therefore, assuming Le = 1 is considered reasonable for methane under the present conditions. Profiles of CO, OH, and $CH_2O$ are additionally compared in Supplementary Fig. S1. Although some quantitative differences are observed, the OpenFOAM calculation using the Correa mechanism with Le = 1 shows overall agreement with the corresponding Cantera results using the same mechanism, including the locations of the major peaks. Cantera results obtained using GRI 3.0 are also included as a reference for the mechanism comparison.

## Grid generation and Quasi-DNS resolution

The computational grids are generated using Pointwise v18.2R1. Prism cells are used near the central axis, and hexahedral cells are used in the rest of the domain. Two main grids are prepared, as shown in Fig. 5: a RANS grid with 747,740 cells and a Quasi-DNS grid with 2,585,352 cells. An axisymmetric model is not suitable because it cannot reproduce circumferential vortex structures. A full 360° model would be ideal but prohibitively expensive. To balance physical fidelity and computational cost, a 30° sector model is adopted. Assuming that the representative scale of vortices at the mixing tube exit is about 10% of the tube diameter, the corresponding azimuthal angle is about 12°. A 30° sector is therefore considered sufficient to capture physically reasonable three-dimensional vortices while keeping the mesh size manageable.

The grid is refined in regions that are critical for mixing and flame stabilization. At the boundary between fuel and air flows inside the mixing tube, the radial grid spacing is set to 50 µm to resolve the shear layer. At the mixing tube outlet, the radial grid spacing is 100 µm to represent the flame structure, and the axial grid spacing near the outlet is about 250 µm. The 30° sector is discretized into 28 cells in the azimuthal direction. The corresponding RANS mesh has a radial grid spacing of 100 µm and an axial grid spacing of 500 µm near the mixing tube outlet. The 30° sector is discretized into 14 cells in the azimuthal direction. The development of these local refinement levels is described below.

During the development of the Quasi-DNS mesh, grid sensitivity was investigated near the mixing tube outlet ($x$ = 0), which is the primary region of interest in the present study, by systematically varying the number of cells. Figure 6 shows the mean and standard deviation (SD) of the temperature, heat release rate, $CH_4$ mass fraction, and axial velocity along the centerline for the different mesh levels. The corresponding results along a radial profile 10 mm downstream of the mixing tube exit are provided in Supplementary Fig. S2. Because the radial location results show trends similar to those along the centerline, the centerline results are discussed below for conciseness. The axial velocity component was examined for this assessment because it is the dominant velocity component in the shear layer and outlet regions relevant to the present analysis.

The Medium mesh has the same grid configuration as the RANS mesh shown in Fig. 5(a). Relative to the Medium mesh, the Coarse mesh was generated by halving the number of divisions in each coordinate direction, resulting in 94,430 cells. The Fine mesh was generated by doubling the number of divisions in each coordinate direction, resulting in 5,981,920 cells. The Fine mesh produced the smallest and most stable standard deviations among the tested mesh levels. However, its computational cost was substantially higher than that of the other meshes. To reduce the computational cost while retaining the resolution required in the regions of interest, the Fine mesh was subsequently modified by reducing the number of cells in regions not central to the present analysis, including the outer portion and downstream region of the combustion chamber. The resulting Refined mesh contained 1,568,252 cells. Although the standard deviations obtained with the Refined mesh were slightly higher than those obtained with the Fine mesh, the mean profiles were comparable. In particular, the standard deviation of the axial velocity was slightly higher with the Refined mesh in regions of large velocity fluctuations. However, the mean velocity profile and the overall distribution of velocity fluctuations remained comparable to those obtained with the Fine mesh. This difference was therefore considered acceptable for the present analysis. The Refined mesh was therefore used as the basis for further mesh development.

It should be noted that the multiresolution results shown in Fig. 6 and Fig. S2 were obtained using an earlier numerical configuration. The geometry, operating condition, chemical mechanism, boundary conditions, and spatial discretization strategy were identical to those used in the final simulations; however, the convergence criteria for the internal iterations were subsequently tightened for the production simulations. The results are therefore presented to document the mesh development and local refinement procedure. Accordingly, this assessment documents the mesh-development process and local grid sensitivity but does not establish formal grid independence of the final production simulation.

Subsequently, the mesh was further refined by reducing the radial grid spacing in the combustion chamber, particularly near the mixing tube outlet, and by reducing the axial grid spacing in the upstream portion of the mixing tube. Because the $CH_4$ mass fraction obtained with the Refined mesh remained slightly higher than that obtained with the Fine mesh from the mixing tube outlet to $x$ = 0.02 m, this difference suggested that further refinement was required to better resolve the mixing process in this region. The mesh was refined iteratively on the basis of these assessments until the final mesh configuration was obtained. The final Quasi-DNS mesh shown in Fig. 5(b), containing 2,585,352 cells, was selected on the basis of this mesh-development procedure.

Based on the mesh development procedure described above, the final Quasi-DNS mesh was assessed using axial and radial distributions of velocity, temperature, and $CH_4$ mass fraction, together with the standard deviations of these quantities along the centerline. In addition, the peak value of the squared mixture fraction gradient near the mixing tube outlet and in the flame anchoring region was used as an additional indicator of local mesh adequacy. In this study, Quasi-DNS denotes a high-resolution, turbulence-model-free calculation for which local mesh sensitivity and flame-structure resolution were assessed in the principal region of interest. The calculation does not satisfy the strict definition of DNS and should not be interpreted as globally grid-independent. Time-averaged fields over a 16 ms period were extracted from the Quasi-DNS solutions for comparison with the unsteady RANS results. The sensitivity of the centerline mean profiles to the averaging duration was assessed using nested 4, 8, and 16 ms intervals ending at the same final sampling time. As shown in Supplementary Fig. S3, the time-averaged temperature, heat release rate, and $CH_4$ mass fraction profiles near the mixing tube outlet were similar among the three averaging intervals. These comparisons support the use of the 16 ms interval for the mean-field comparisons in the present stable lean case, while assessing sensitivity to averaging duration rather than establishing convergence of all temporal statistics. Statistical convergence of the temporal statistics was not demonstrated; comparisons based on the 16 ms window, particularly for the heat release rate, should therefore be interpreted cautiously.

**Diagnostics: mixture fraction gradient and flame index**

To characterize mixing and combustion modes near the mixing tube outlet, two diagnostics are employed: the mixture fraction gradient and flame index [42, 44-45]. The squared magnitude of the mixture fraction gradient is defined as

$$|\nabla Z|^2 = \left(\frac{\partial Z}{\partial x}\right)^2 + \left(\frac{\partial Z}{\partial r}\right)^2 + \left(\frac{1}{r}\frac{\partial Z}{\partial \theta}\right)^2 \qquad (1)$$

where $Z$ is the mixture fraction, and $x$, $r$, and $\theta$ denote the axial, radial, and azimuthal coordinates, respectively. This quantity characterizes the local mixing layer structure in the shear layer and is used to examine its spatial relation to the flame anchoring region.

The combustion mode is evaluated using the flame index based on methane. The flame index is

$$FI = \frac{1}{2}\left(1 + \frac{\nabla Y_{CH4} \cdot \nabla Y_{O2}}{|\nabla Y_{CH4}||\nabla Y_{O2}|}\right) \qquad (2)$$

Here, $\nabla Y_{CH4}$ and $\nabla Y_{O2}$ are the methane and oxygen mass fractions. Values of FI > 0.5 indicate premixed-like combustion, whereas values below 0.5 indicate diffusion-dominated combustion. To focus the analysis on regions associated with appreciable chemical heat release, flame index contours are shown only in regions where the heat release rate, $q$, exceeds 1% of the maximum heat release rate within each evaluation region, $q_{max}$. For visualization purposes, the flame index is set to 0.5 in regions where the heat release rate is below this threshold. For selected cases with low heat release rates, contours based on a threshold of 0.1% of $q_{max}$ are additionally presented to visualize the flame index near the mixing tube outlet. In this workflow, the filtered flame index contours are analyzed near the mixing tube outlet to classify the local combustion state and to compare how RANS and Quasi-DNS represent premixed versus diffusion-like regions under partially premixed conditions.

## Comparison of Quasi-DNS and RANS results

The Quasi-DNS workflow is compared with unsteady RANS solutions for a representative lean partially premixed case and with a perfectly premixed reference case at the same overall equivalence ratio. The goal is to show that, for this mixing tube burner element, the method captures the flame structure and stabilization mechanism near the mixing tube outlet differently from the corresponding RANS solution under conditions relevant to the multiple-injection burner. The RANS results are used as an engineering level reference rather than as validation of the Quasi-DNS approach; a systematic comparison with LES is beyond the scope of the present study.

### Partially premixed flame at $\phi = 0.45$

The primary case considered here is a partially premixed methane flame at an overall equivalence ratio of $\phi = 0.45$, using the coaxial inlet and thermal conditions described in the Method details. Figure 7 compares contours of temperature, heat-release rate, $CH_4$ mass fraction and absolute velocity for this case, showing RANS, instantaneous Quasi-DNS and time-averaged Quasi-DNS fields. In the present workflow, the flame lift-off height $H_L$ is defined as the axial position where the OH mass fraction first exceeds 0.0002, following previous studies of turbulent lifted flames in hot vitiated coflow [43]. The lift-off position is indicated by a dotted line in the contour plots, and $H_L$ denotes this lift-off height. Both approaches produce a flame stabilized near the mixing tube outlet, but the resolved structures differ markedly.

In the RANS calculation, the flame appears as a relatively smooth, axisymmetric front anchored very close to the tube exit. Because the k-ε model and fine grid promote rapid turbulent mixing, the fuel-air mixture is nearly uniform at the outlet, and the flame resembles a premixed flame with a short lift-off height and a compact heat release region. In the Quasi-DNS results, fuel-air mixing inside the mixing tube remains limited, and distinct vortical structures form at the tube outlet. These vortices enhance mixing locally and produce an extended three-dimensional heat release region downstream of the exit, with a slightly larger lift-off height and localized high heat release pockets near the shear layer.

Figure 8 focuses on the mixing tube outlet region and shows contours of squared mixture fraction gradient magnitude and its axial and radial components, together with heat release rate, for the Quasi-DNS solution at $\phi = 0.45$. In the flame anchoring region, the peak squared mixture fraction gradient magnitude is on the order of $10^3$ $m^{-2}$ and is confined to a narrow band near the shear layer. Figure 9 presents counterflow diffusion flames computed with Cantera for several stretch rates and mesh resolutions. For low-stretch cases with peak $|\nabla Z|^2$ values comparable to those in the flame anchoring region shown in Fig. 8, differences in the temperature profile between the adaptive mesh and uniform grids with spacings of 50-250 μm remain small. Taken together, these results support the selected 50 μm radial and 250 μm axial grid spacings for resolving the flame structure near the mixing tube outlet.

The combustion mode is examined using the flame index. Figure 10(a) and (b) show the flame index and heat release rate for the RANS and Quasi-DNS solutions, respectively, of the partially premixed flame at $\phi = 0.45$. Because the heat release rate near the mixing tube outlet is low in the Quasi-DNS result, an additional flame index contour using a heat release rate threshold of 0.1% of $q_{max}$ is shown on the right in Fig. 10(b). Although the flame index values differ between the two results, the flame index is greater than 0.5 near the mixing tube outlet in both cases, indicating premixed-like combustion. Locally diffusion dominated combustion is also observed in the downstream portion of the heat release region, where the computational mesh is relatively coarse.

### Perfectly premixed reference at $\phi = 0.45$

To provide a premixed reference for interpreting the partially premixed Quasi-DNS result, a perfectly premixed reference case at $\phi$ = 0.45 is simulated by supplying a homogeneous methane-air mixture at the inlet, while keeping the same geometry, flow rate and thermal boundary conditions. Figure 11 presents Quasi-DNS results for this perfectly premixed flame, including temperature, heat release rate, $CH_4$ mass fraction and velocity fields near the mixing tube outlet. Figure 12 compares the axial distributions of integrated heat release rate at $\phi$ = 0.45 for RANS, the partially premixed Quasi-DNS case and the perfectly premixed Quasi-DNS reference. In the perfectly premixed configuration, the axial location and thickness of the main heat release region predicted by Quasi-DNS are close to those of the RANS solution for the partially premixed configuration. The flame is attached to the mixing-tube outlet ($H_L \simeq 0$), as shown in Fig. 11, and provides a premixed reference for interpreting the downstream shift and broadening of the heat release distribution in the partially premixed Quasi-DNS case.

In the partially premixed case, however, Figure 12 shows that Quasi-DNS predicts a downstream shift and broadening of the heat-release distribution compared with RANS, reflecting the limited mixing in the mixing tube and the role of vortices at the outlet. As expected, the flame index in Fig. 10(c) is greater than 0.5 throughout most of the heat release region for the perfectly premixed Quasi-DNS case, although isolated diffusion dominated regions are observed. Taken together, the comparisons in Figures 7, 8, 10, 11 and 12 show the Quasi-DNS results for a reference premixed flame and, at the same time, provide additional detail on mixing, flame structure and combustion mode in the partially premixed case that is not resolved by RANS. The present comparison is limited to stable lean conditions at $\phi$ = 0.45 and a perfectly premixed reference at the same equivalence ratio.

## Limitations

The Quasi-DNS workflow presented in this study has several limitations that should be considered when applying it to other burner designs or operating conditions.

First, the method is developed and applied for a single burner element with a simplified geometry. The coaxial inlet and mixing tube are modeled with zero thickness partitions and do not include features such as tube wall thickness, cooling holes or structural supports that exist in practical multiple-injection burners. These simplifications allow the essential shear layer mixing and flame anchoring mechanisms near the mixing tube outlet to be resolved at a feasible computational cost but limit the direct quantitative applicability of the results to full combustor assemblies. The present results are therefore best interpreted at the burner element scale and as qualitative guidance for scaling trends in more complex hardware.

Second, all simulations are performed at atmospheric pressure with preheated air at 400 °C and methane as the only fuel. Under practical gas turbine conditions, higher pressure and density change laminar burning velocity, ignition delay, and Reynolds number, leading to thinner flames that would require finer grids and, in most cases, a combustion model. The present workflow should therefore be regarded as providing qualitative insight into flame structure and stabilization rather than quantitative predictions for high pressure engines.

Third, the chemical and transport models are simplified. The Correa mechanism is an efficient choice for methane-air combustion under the present conditions, as assessed by comparison with GRI 3.0 and available experimental data, and the assumption of Le = 1 is acceptable for atmospheric methane, as checked using Cantera-based reference calculations. However, additional validation and possibly extended mechanisms would be required for fuel blends containing hydrogen or heavier hydrocarbons, and for conditions where preferential diffusion plays a significant role.

In this sense, the present Quasi-DNS calculations should be regarded as high-resolution simulations performed without an explicit turbulence model or subgrid scale combustion model. They resolve flame structures and key vortical features near the mixing tube outlet, but do not resolve the full range of turbulent scales at practical engine Reynolds numbers and therefore do not satisfy the strict definition of DNS.

Finally, the Quasi-DNS simulations are still computationally demanding and do not resolve all turbulent scales at high Reynolds numbers. The three-dimensional grids contain 747,740 cells for RANS and 2,585,352 cells for Quasi-DNS, and time-resolved runs must cover tens of milliseconds of physical time to obtain statistically meaningful averages. As a result, the workflow is best suited for targeted analyses of flame structure and stabilization in coaxial burners with mixing tubes and for assessing the limitations of RANS-based design calculations. It is not intended for routine parametric design studies in full scale gas turbine combustors. Despite these limitations, the workflow is directly reusable for single element studies of mixing tube burners and for benchmarking and improving RANS-based combustion models under lean, stable operating conditions. The method is intended primarily for researchers who require a high-resolution numerical reference under the stated assumptions for comparisons with RANS or LES combustion models in mixing tube type gas turbine burner elements. Quantitative accuracy for the physical burner has not been established. The present study is limited to a stable lean operating condition at $\phi = 0.45$ and does not assess extinction limits, transient lean blow-out events, or the predictive capability of the present approach for lean blow-out.

## Ethics statements

N/A

## CRediT author statement

**Kazuki Abe:** Conceptualization, Methodology, Software, Investigation, Visualization, Writing – original draft. **Youhi Morii:** Conceptualization, Writing – review & editing, Supervision. **Kaoru Maruta:** Writing – review & editing, Supervision.

## Acknowledgments

This research did not receive any specific grant from funding agencies in the public, commercial, or not-for-profit sectors.

## Declaration of interests

☒ The authors declare that they have no known competing financial interests or personal relationships that could have appeared to influence the work reported in this paper.

☐ The authors declare the following financial interests/personal relationships which may be considered as potential competing interests:

## Supplementary material *and/or* additional information [OPTIONAL]

Supplementary material associated with this article includes comparisons of species profiles for the counterflow diffusion flames (Fig. S1), radial profiles used in the mesh-sensitivity assessment (Fig. S2), and the sensitivity of the centerline time-averaged Quasi-DNS profiles to the averaging duration (Fig. S3).

## Declaration of Generative AI and AI-assisted technologies in the writing process

An AI-assisted writing tool was used only to improve the language and readability of the manuscript. After using this tool, the authors reviewed and edited the content as needed and take full responsibility for the content of the publication.

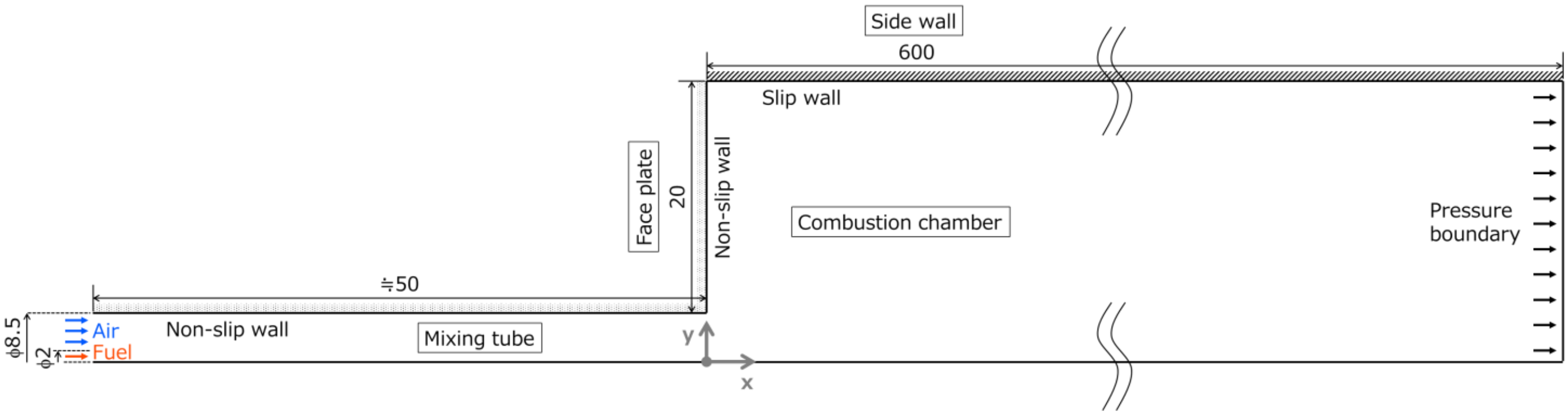

Fig. 1 Longitudinal section of a 30° computational sector of a single multiple-injection burner element. A mixing tube is positioned downstream of the coaxial fuel-air nozzle, and combustion occurs in the downstream combustion chamber. Periodic boundary conditions are applied on the azimuthal sector faces. Dimensions are in millimeters.

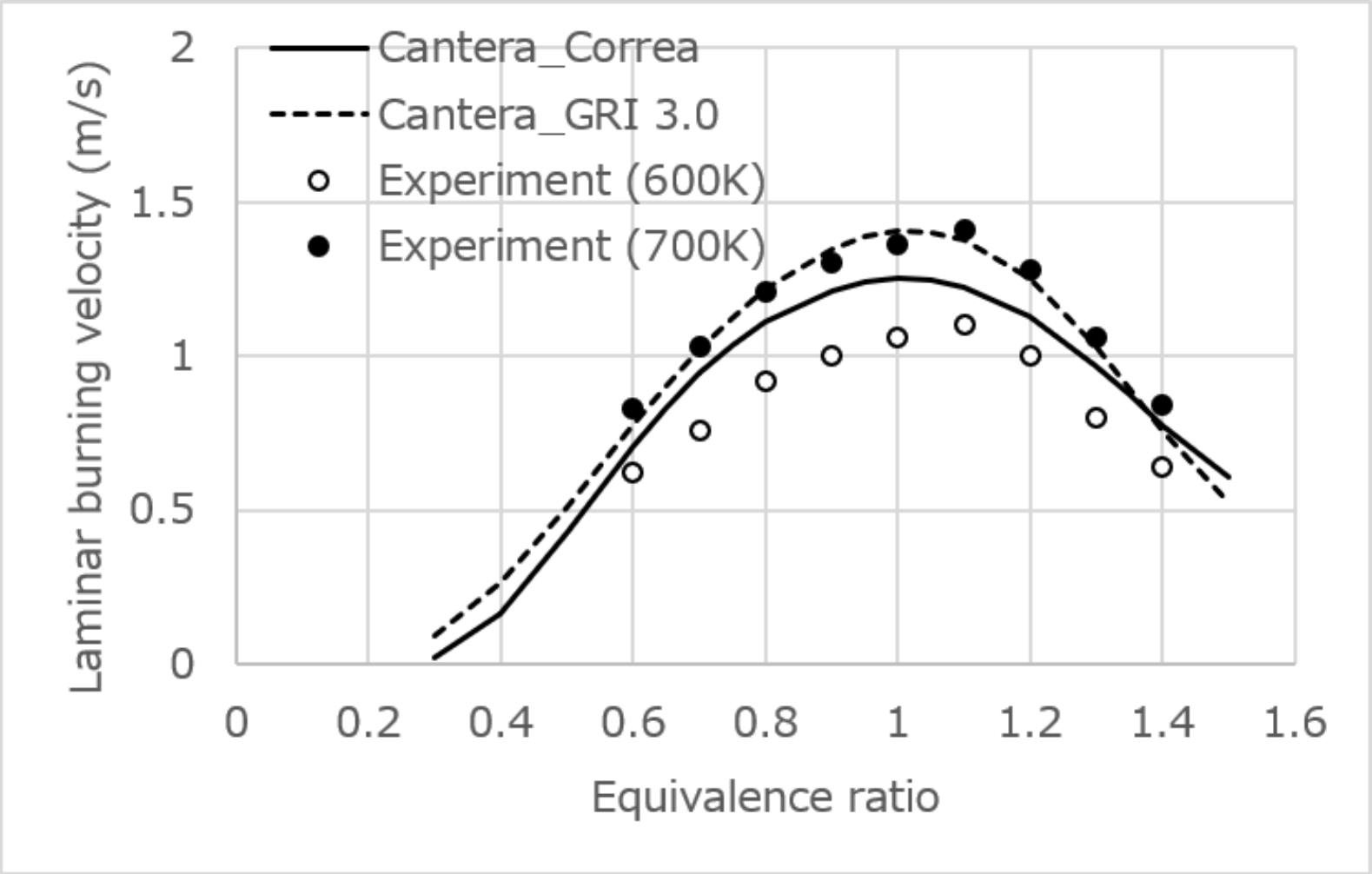

Fig. 2 Comparison of laminar burning velocities between two mechanisms: Correa (solid line) and GRI 3.0 (dotted line). Experimental data from Ref. [46] are included for comparison. Air and fuel are at 400 °C and 25 °C, respectively, at atmospheric pressure.

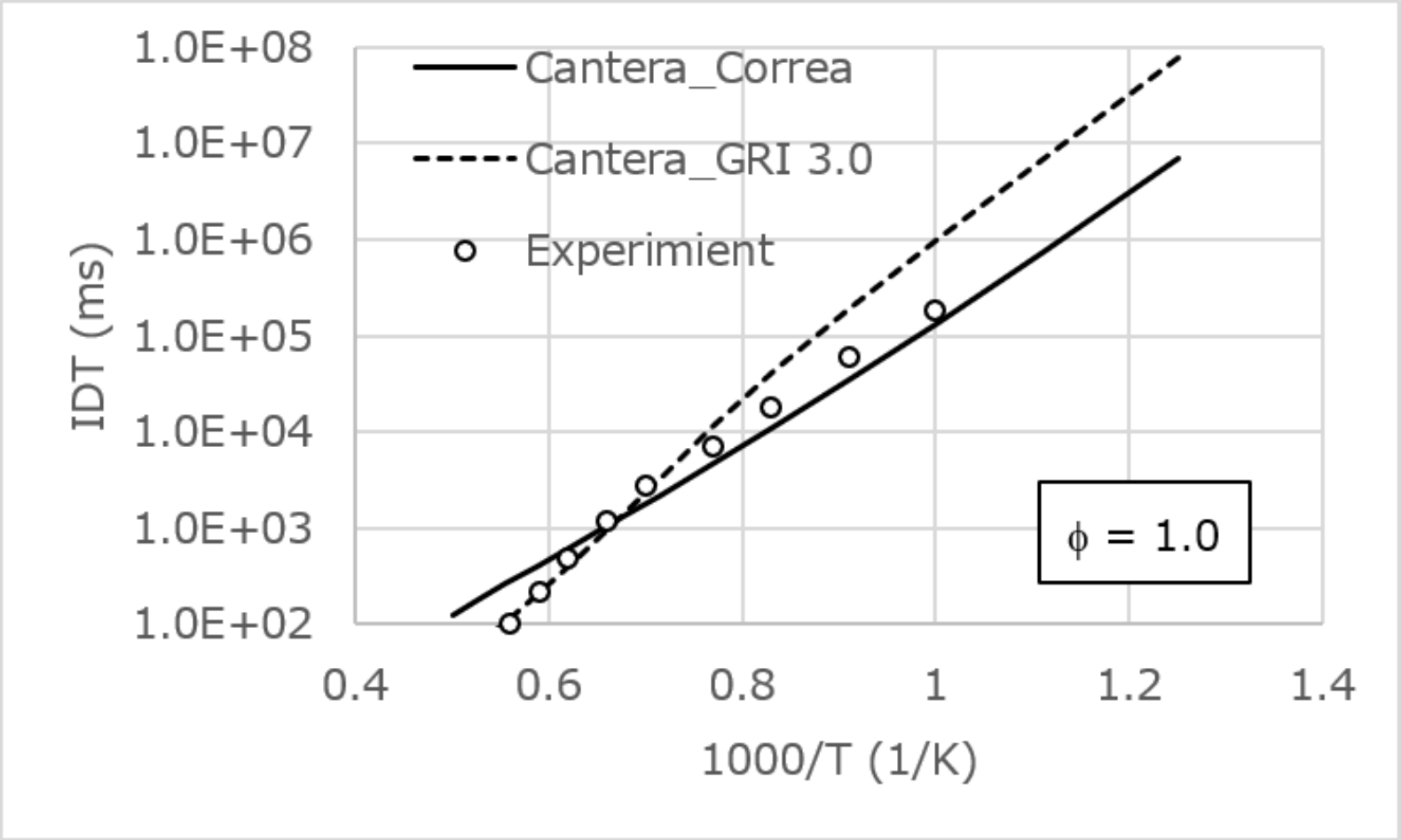

Fig. 3 Comparison of ignition delays between two mechanisms at atmospheric pressure: Correa (solid line) and GRI 3.0 (dotted line). Results are shown for an equivalence ratio of ϕ = 1.0. Experimental data from Ref. [47] are included for comparison.

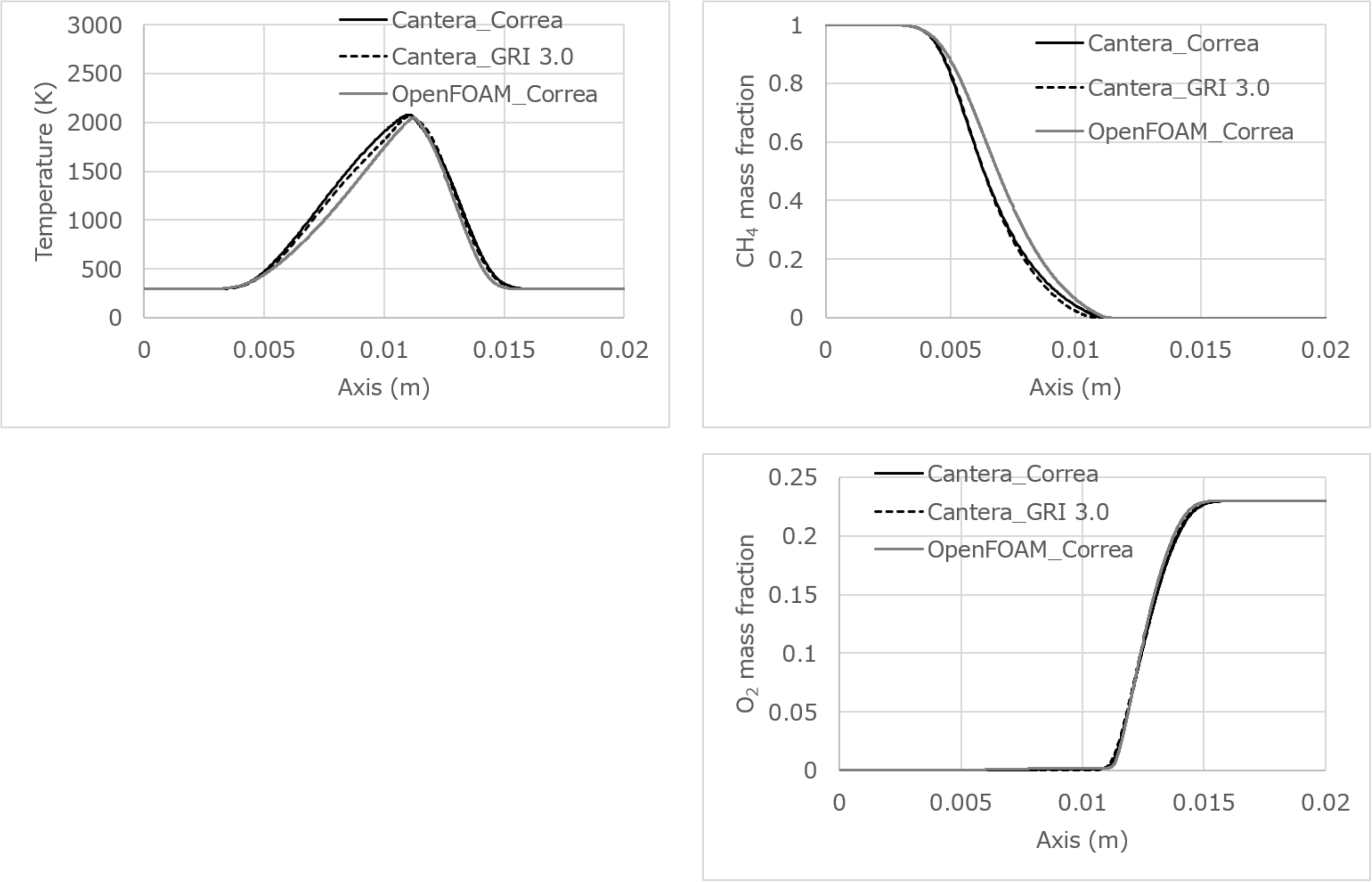

Fig. 4 Comparison of counterflow diffusion flames at atmospheric pressure with air and fuel at 25 °C. Temperature, $CH_4$ mass fraction and $O_2$ mass fraction profiles are shown for Cantera with the Correa mechanism (black solid line), Cantera with GRI 3.0 (black dotted line) and OpenFOAM (reactingFoam) with the Correa mechanism (grey solid line). The stretch rate is 11.7 $s^{-1}$.

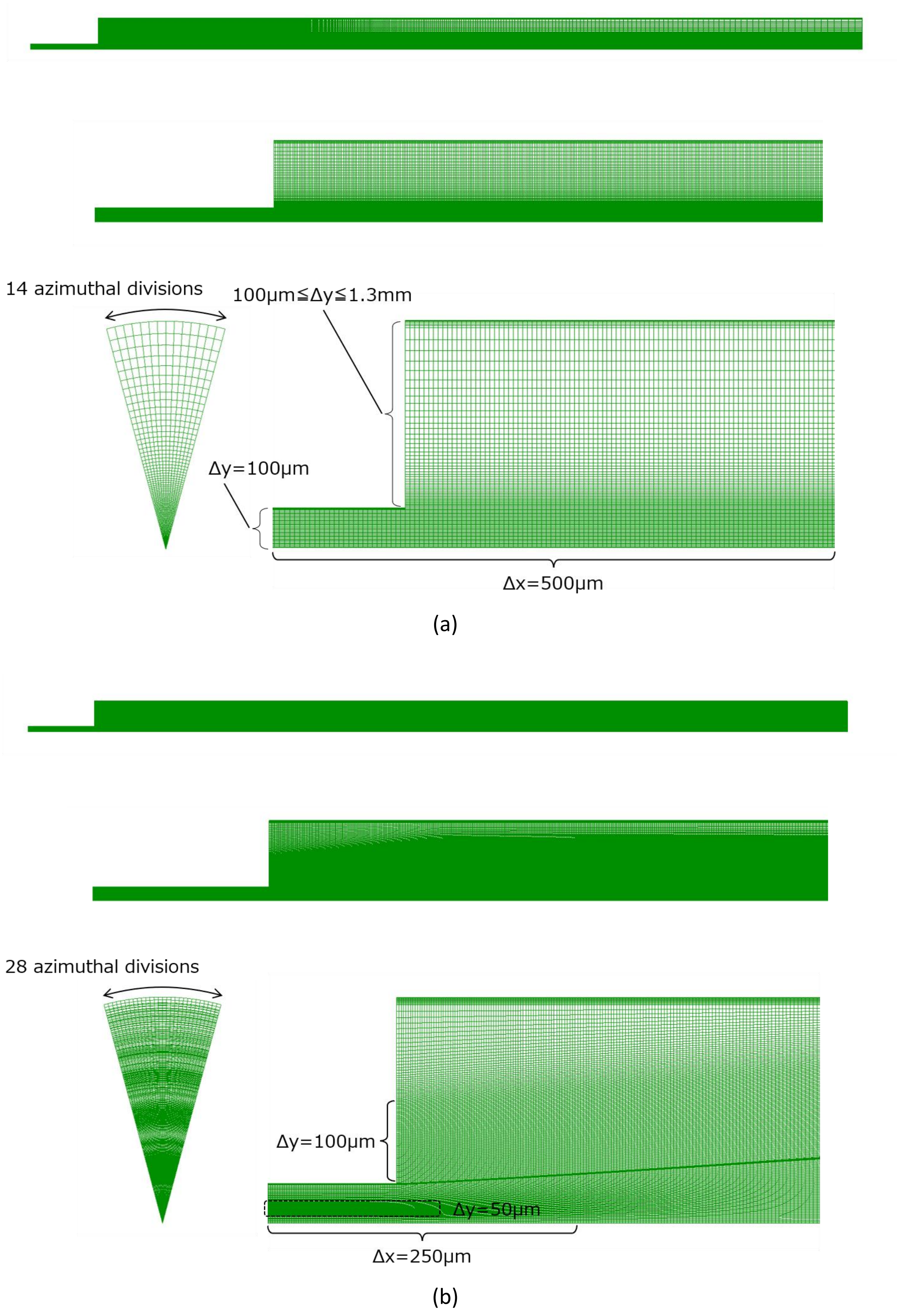

Fig. 5 Grid structure of (a) the RANS calculation with 747,740 cells and (b) the Quasi-DNS calculation with 2,585,352 cells. The Quasi-DNS grid has a radial spacing of 50 μm inside the mixing tube and 100 μm near the outer periphery, and an axial spacing of 250 μm near the mixing tube outlet.

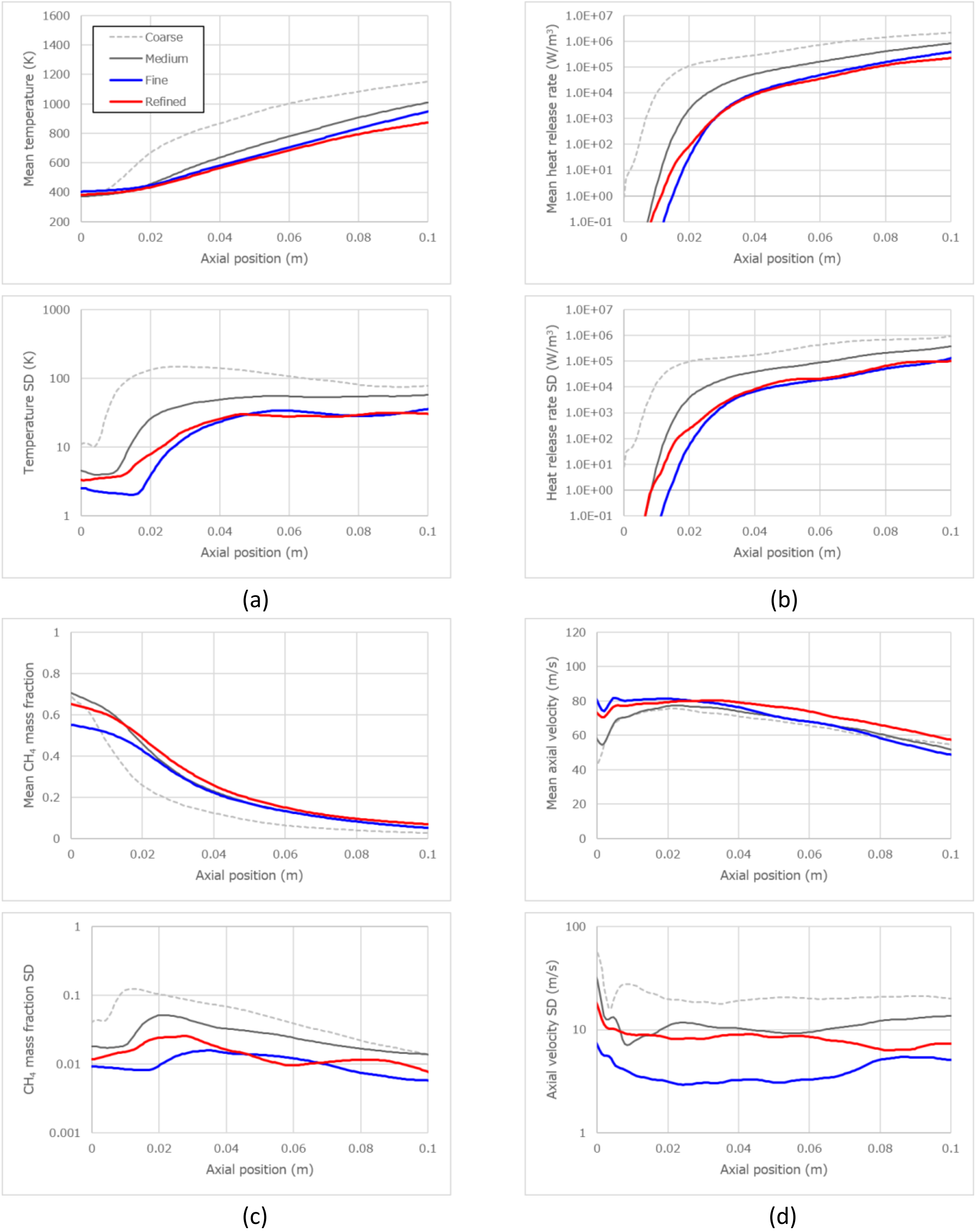

Fig. 6 Effects of mesh resolution on the mean and standard deviation (SD) of temperature, heat release rate, $CH_4$ mass fraction, and axial velocity along the centerline near the mixing tube outlet. The mixing tube outlet is located at $x = 0$.

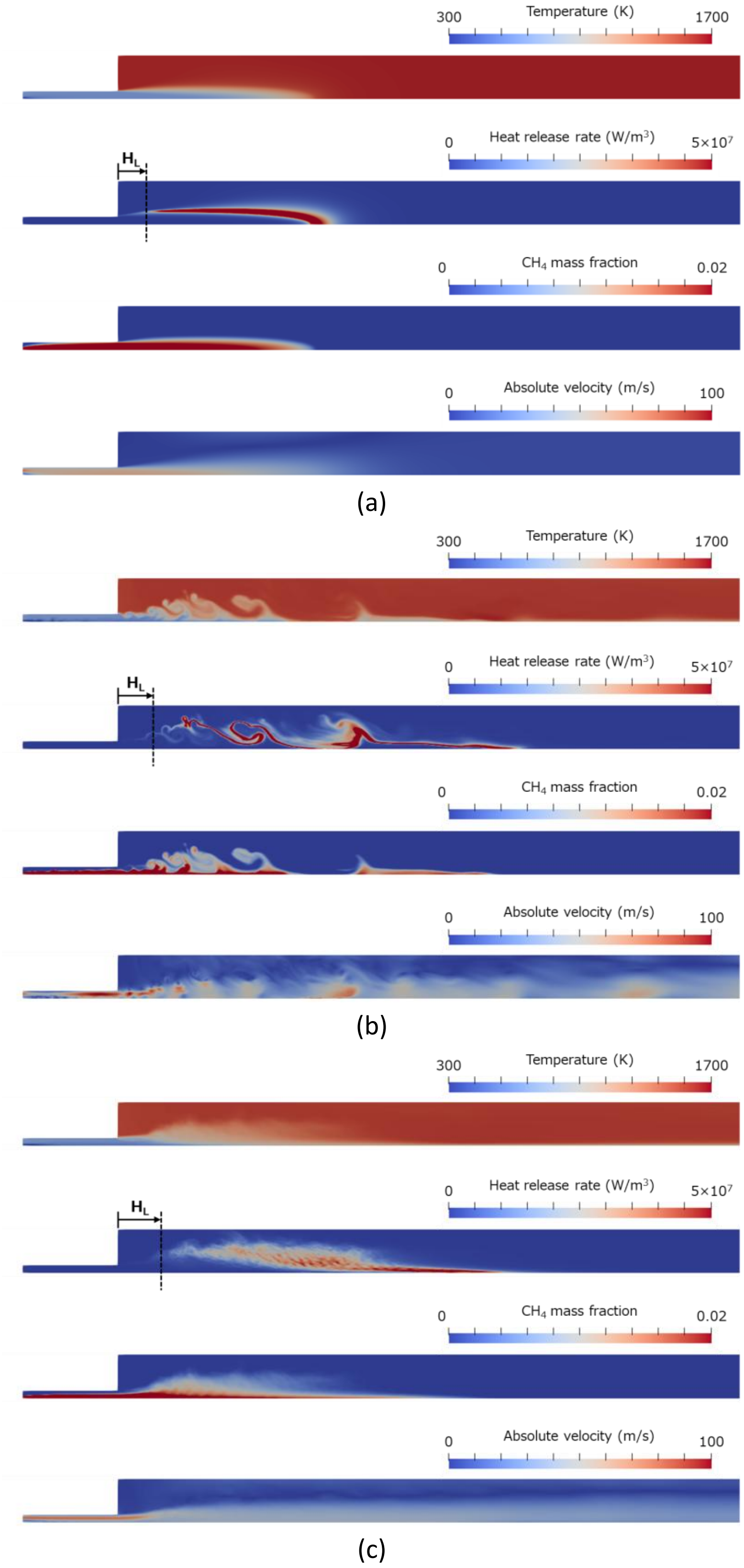

Fig. 7 Comparison of simulated fields under stable conditions at $\phi = 0.45$. Distributions of temperature, heat release rate, $CH_4$ mass fraction and absolute velocity are shown for (a) RANS, (b) instantaneous Quasi-DNS, and (c) time-averaged Quasi-DNS.

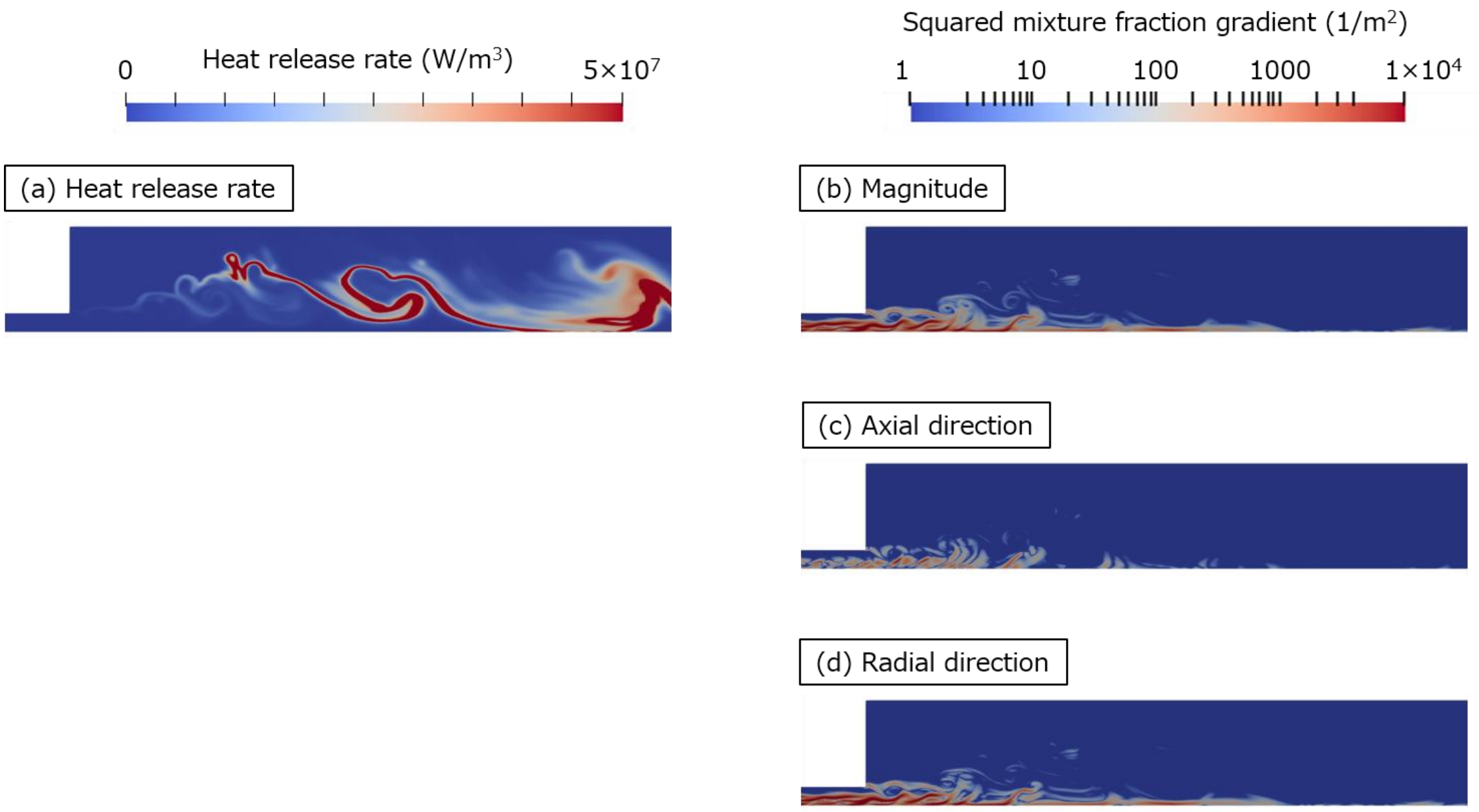

Fig. 8 Quasi-DNS instantaneous contours of (a) heat release rate, (b) the squared mixture fraction gradient magnitude, $|\nabla Z|^2$, (c) its axial component, $(\partial Z/\partial x)^2$, and (d) its radial component, $(\partial Z/\partial r)^2$, focusing on the flame anchoring region at $\phi$ = 0.45.

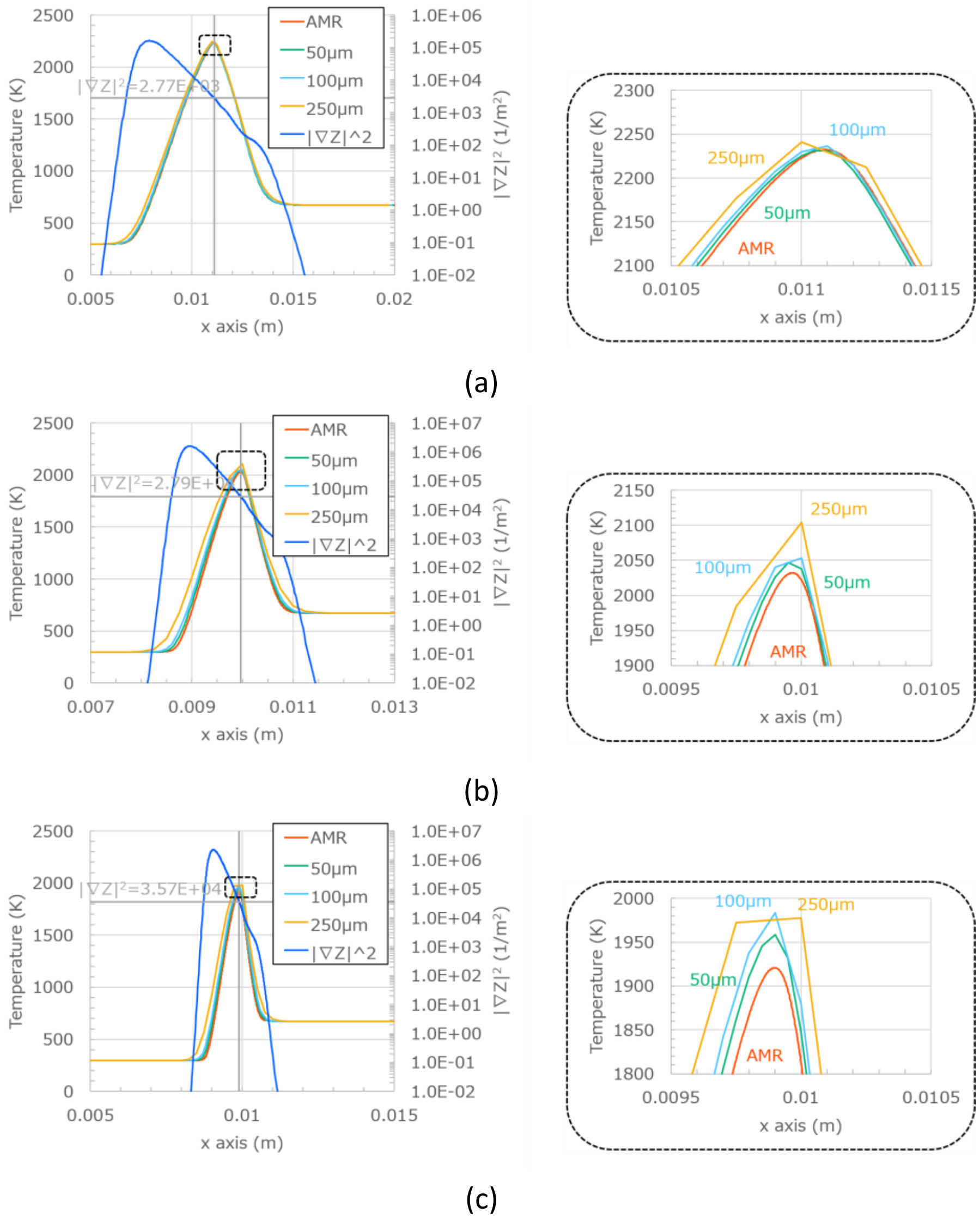

Fig. 9 Results of counterflow diffusion flames computed using Cantera to assess the adequacy of the grid spacings used in the Quasi-DNS calculations shown in Fig. 7(b), (c). The stretch rates and peak $|\nabla Z|^2$ values at highest temperature positions are (a) 65.1 $s^{-1}$ and $2.77\times10^3$ $m^{-2}$ , (b) 607.3 $s^{-1}$ and $2.79\times10^4$ $m^{-2}$ , (c) 799.4 $s^{-1}$ and $3.57\times10^4$ $m^{-2}$, respectively.

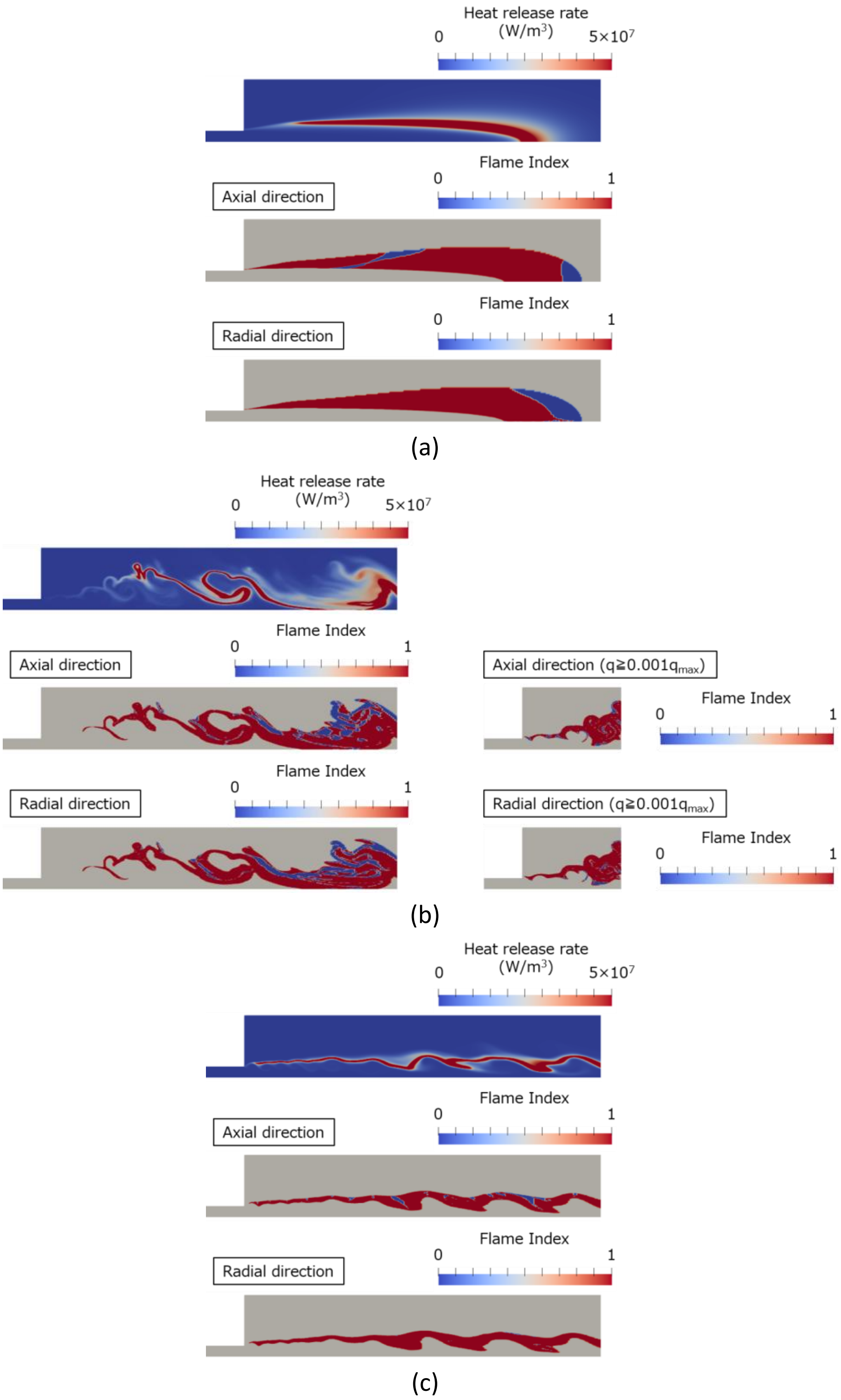

Fig. 10 Flame index fields calculated from Eq. (2) at ϕ = 0.45: (a) RANS, (b) partially premixed Quasi-DNS, and (c) perfectly premixed Quasi-DNS, together with the heat release rate field. Regions with heat release rates exceeding 1% of the maximum heat release rate, $q_{max}$, are shown. For (b), a contour plot showing regions exceeding 0.1% of $q_{max}$ is additionally presented on the right to examine the flame index near the mixing tube exit.

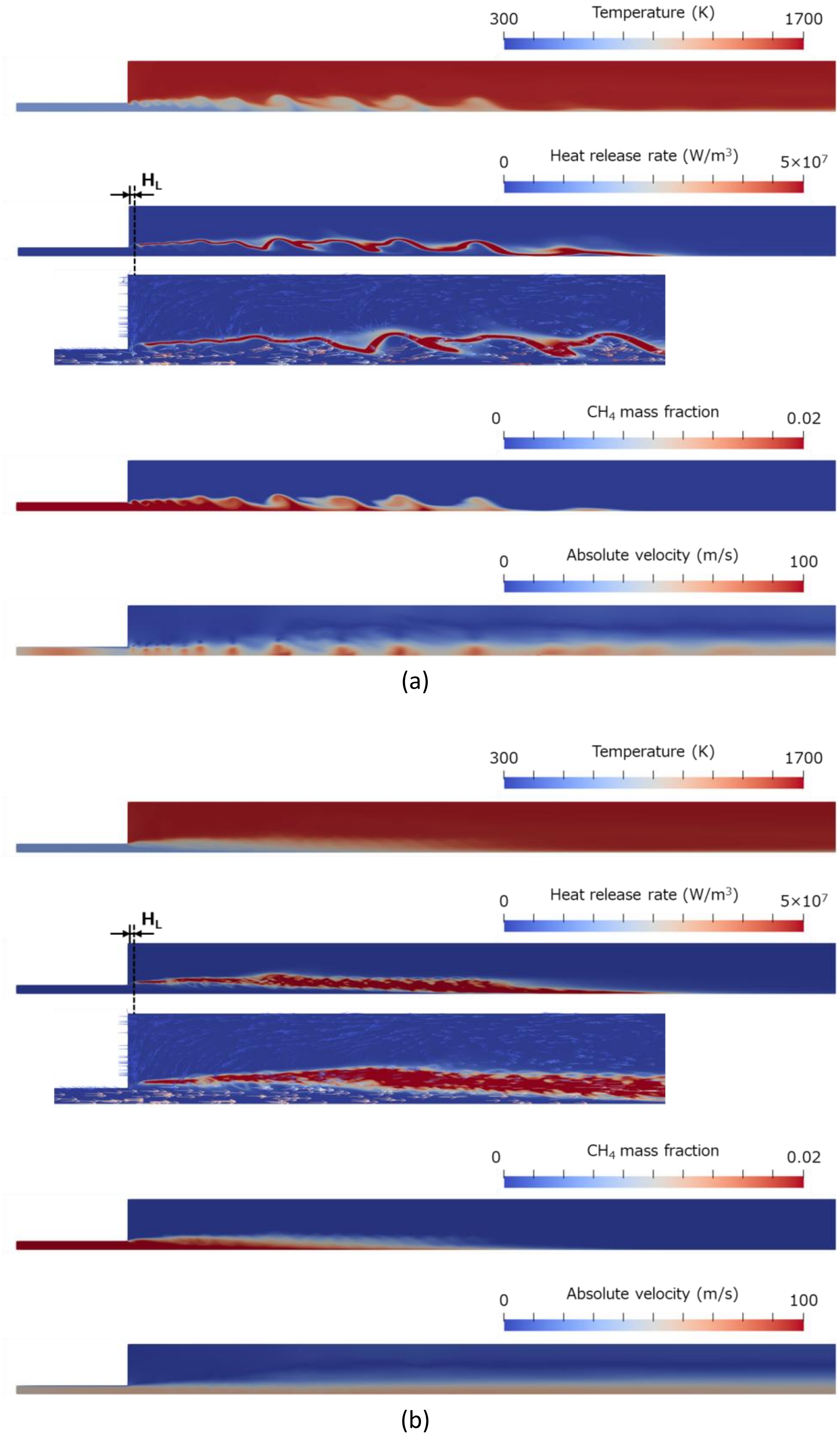

Fig. 11 Quasi-DNS results for the perfectly premixed reference flame at $\phi = 0.45$: (a) instantaneous fields and (b) time-averaged fields. The flame is attached to the mixing tube outlet (lift-off height $H_L \simeq 0$).

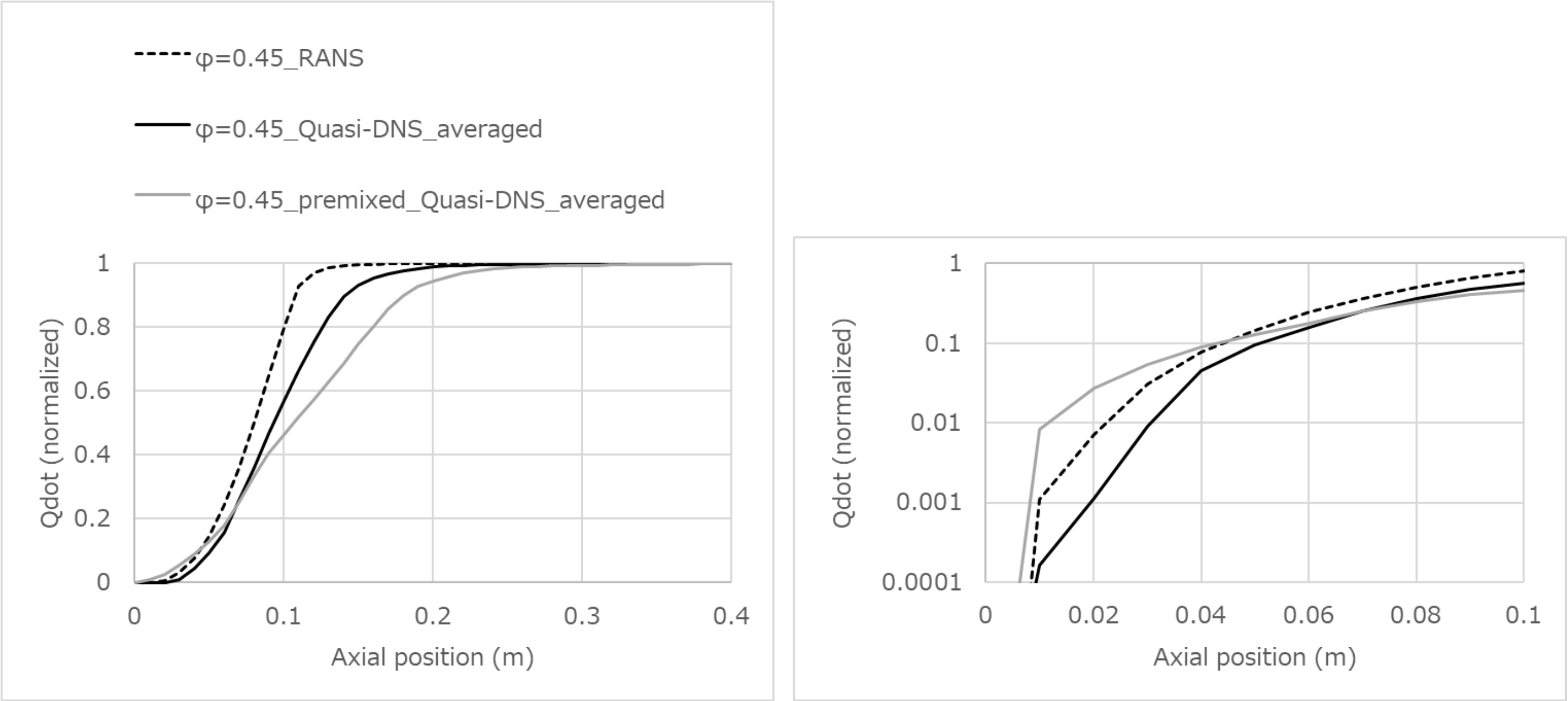

Fig. 12 Axial profiles of integrated heat release rate at ϕ = 0.45. Shown are RANS, partially premixed Quasi-DNS, and perfectly premixed Quasi-DNS results. The Quasi-DNS curves are time-averaged.

Table 1 Calculation conditions: The fuel flow rate is varied according to the set equivalence ratio, while the air flow rate is fixed. The fuel is pure methane. The temperature conditions are equivalent to those of gas turbine combustor operation, and the pressure conditions are atmospheric pressure.

| | Fuel | Air |
|---|---|---|
| Inlet velocity (m/s) | 18.81 (depends on ϕ) | 52.93 |
| Compositions (Mass fractions) | $CH_4$: 1.0 | $O_2$: 0.232<br>$N_2$: 0.768 |
| Temperature (°C) | 25 | 400 |
| Pressure (MPa) | 0.101325 | |
| Equivalence ratio ϕ [a] | 0.45 | |

a. Equivalence ratio ϕ when fuel and air are completely mixed inside the mixing tube.

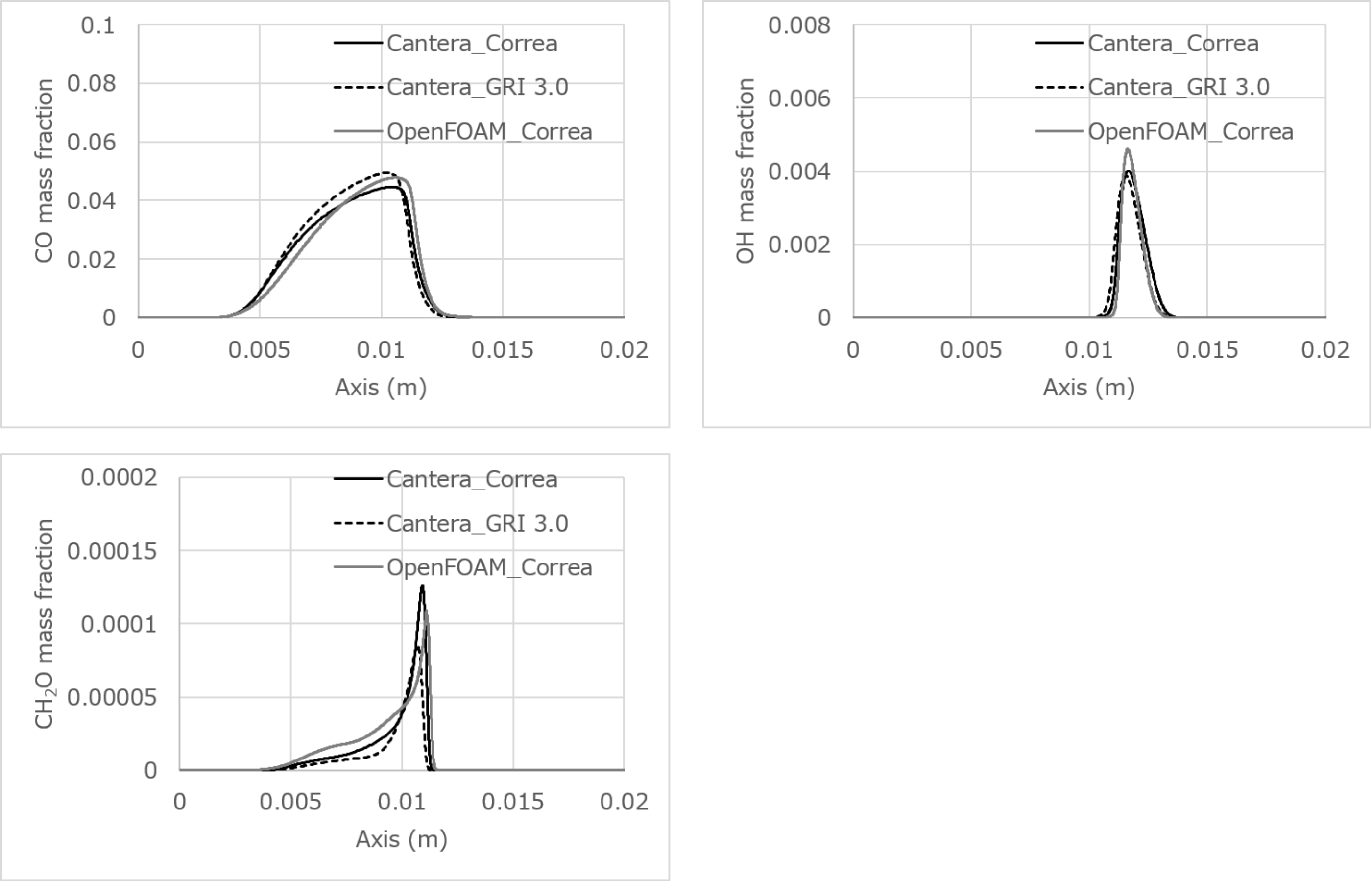

Fig. S1 Profiles of representative intermediate and radical species under the same conditions as those in Fig. 4: (a) CO, (b) OH, and (c) $CH_2O$. Results obtained using OpenFOAM with the Correa mechanism and Le = 1 are compared with those obtained using Cantera with the Correa mechanism and GRI 3.0.

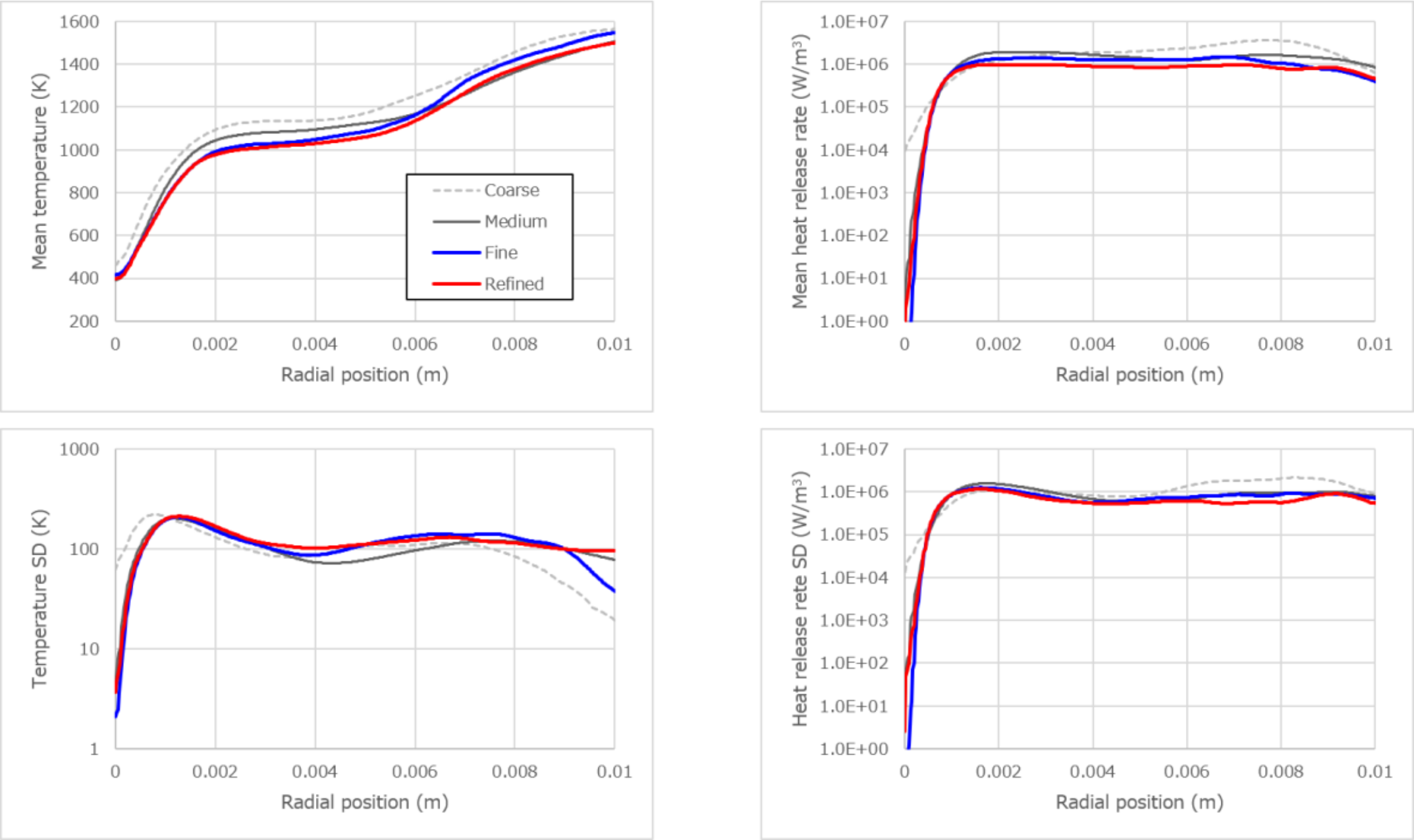

(a) (b)

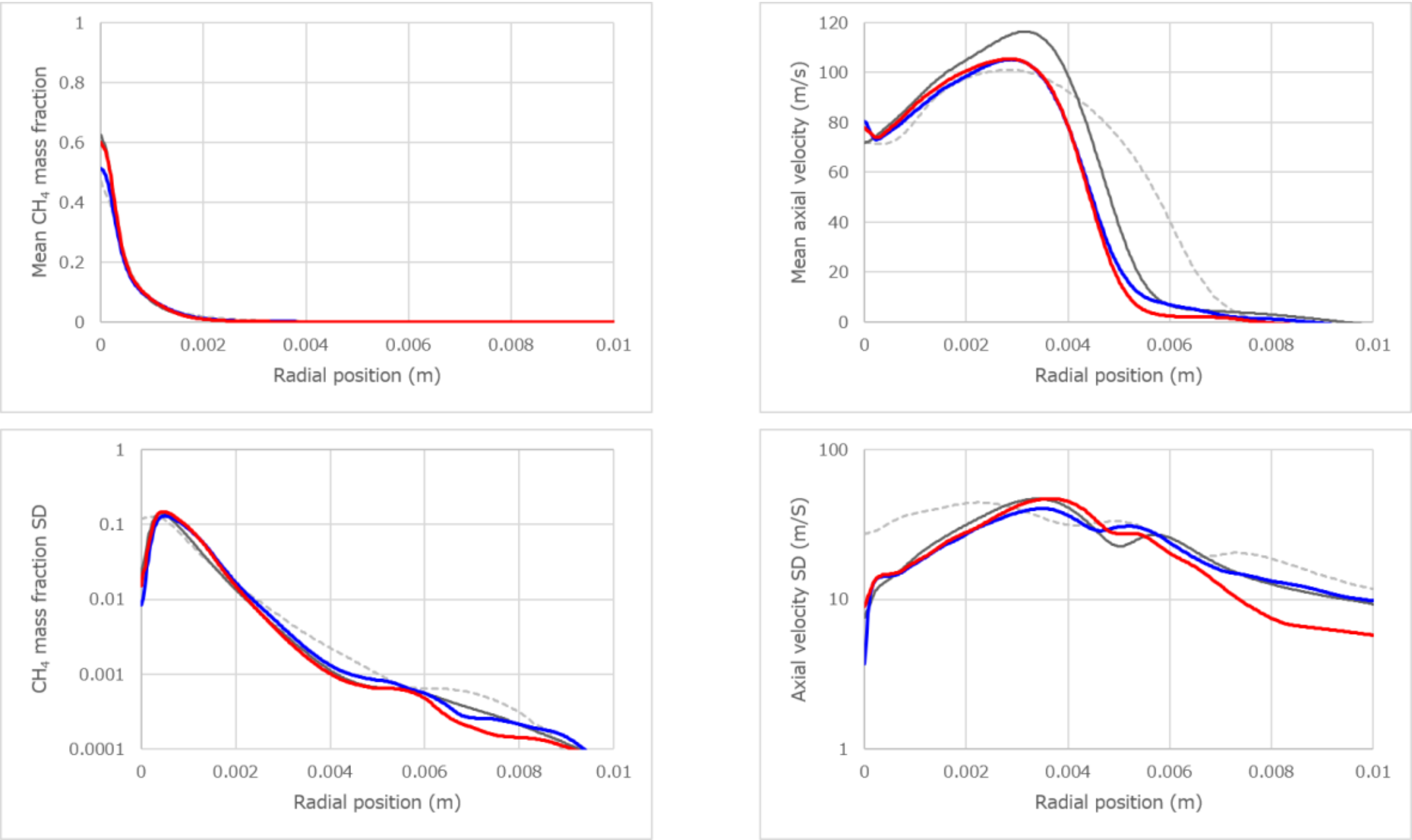

(c) (d)

Fig. S2 Effects of mesh resolution on the mean and standard deviation of the temperature, heat release rate, $CH_4$ mass fraction, and axial velocity along a radial profile 10 mm downstream of the mixing tube exit. The trends are similar to those observed along the centerline (Fig. 6).

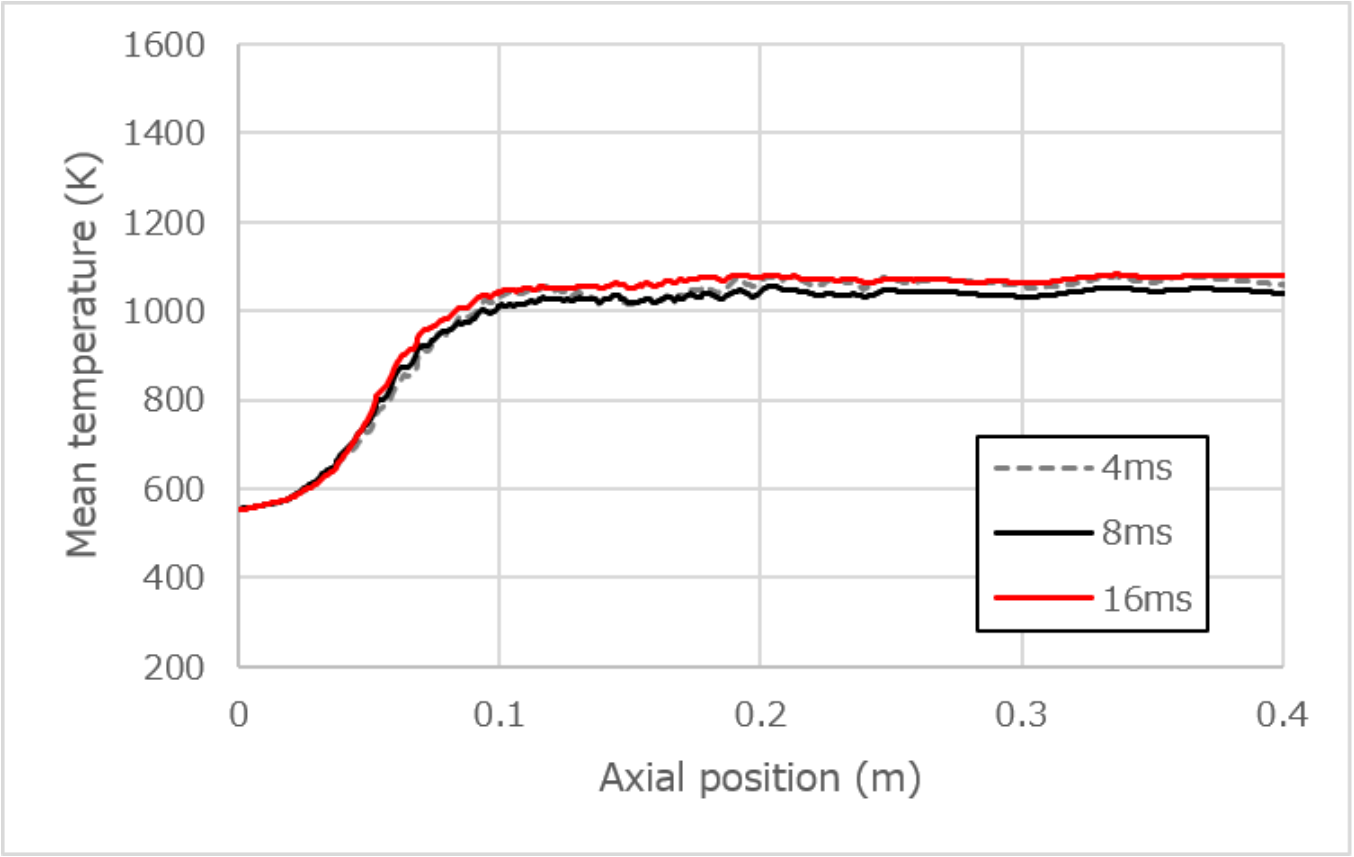

(a)

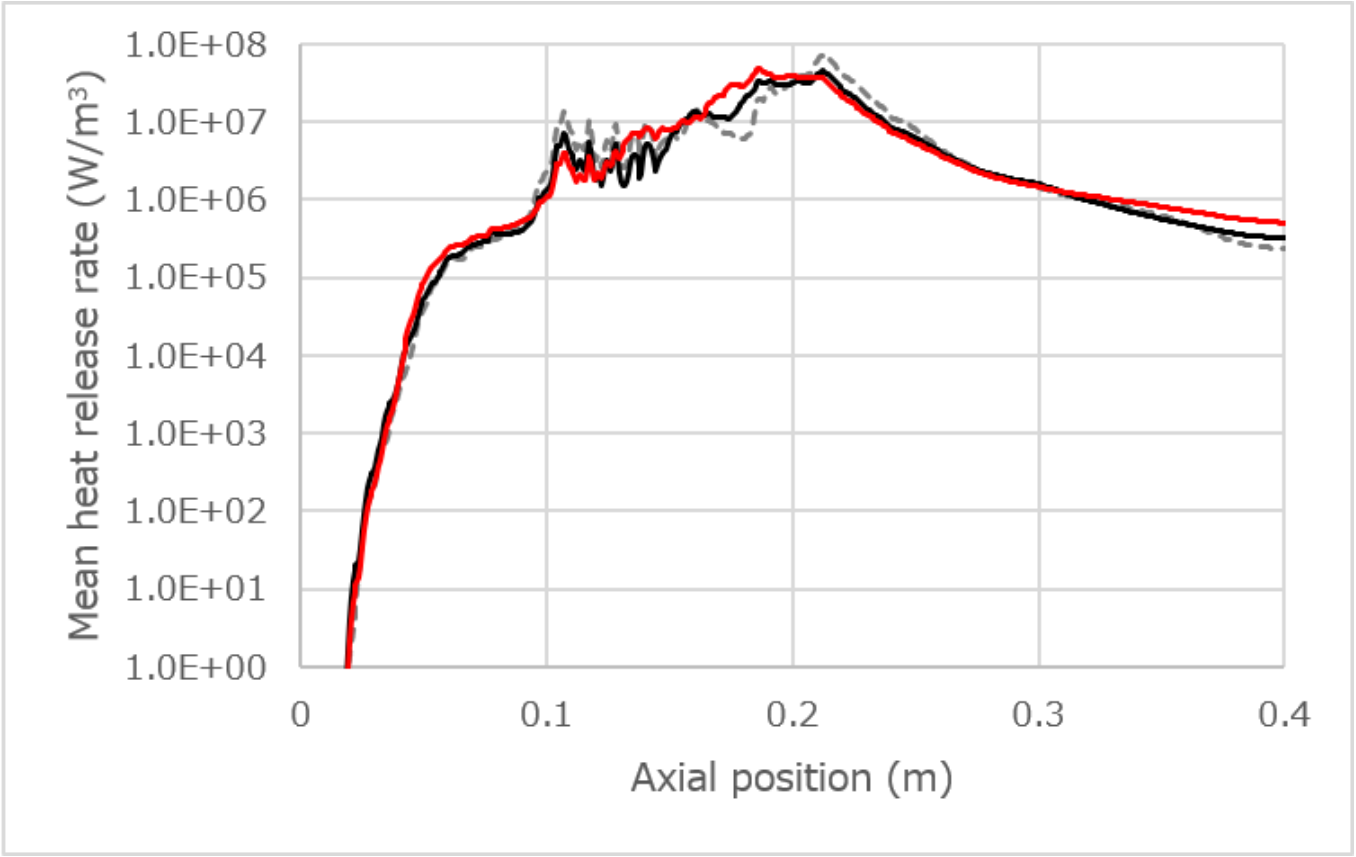

(b)

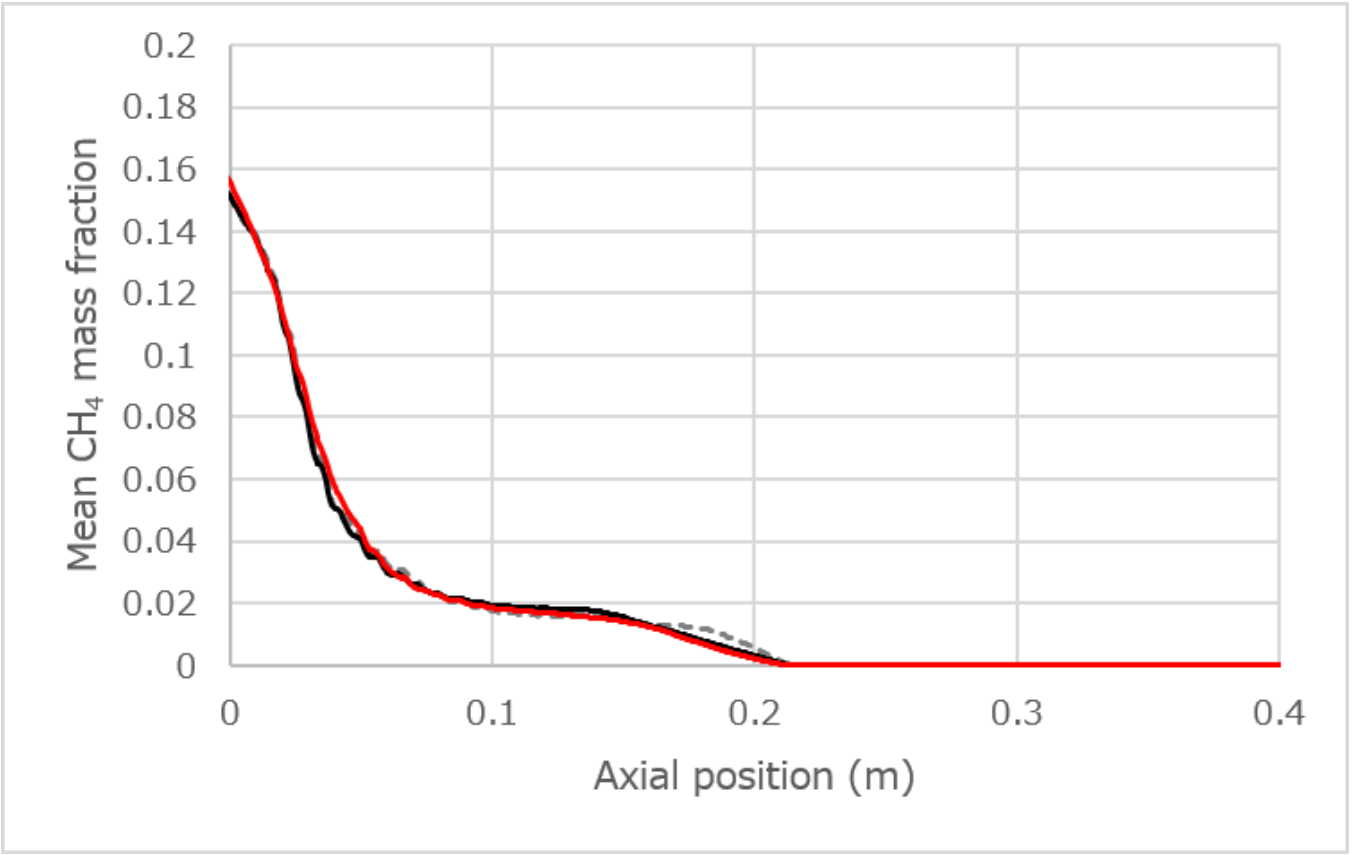

(c)

Fig. S3 Sensitivity of the centerline time-averaged Quasi-DNS profiles to the averaging duration for the partially premixed case at $\phi$ = 0.45. Profiles of (a) temperature, (b) heat release rate, and (c) methane mass fraction were averaged over nested 4 ms, 8 ms, and 16 ms intervals ending at the same final sampling time. The mixing-tube outlet is located at $x$ = 0.